\documentclass[twocolumn,showpacs,preprintnumbers,amsmath,amssymb]{revtex4}

\usepackage{graphicx}% Include figure files
\usepackage{dcolumn}% Align table columns on decimal point
\usepackage{bm}% bold math
\usepackage{epstopdf}
\begin{document}

\preprint{APS/123-QED}

\title{The Role of 
Nonlinear Friction in the Dewetting of Thin Film Polymers}
\author{Thomas Vilmin}
\author{Elie Rapha{\"e}l}
\affiliation{
Laboratoire de Physique des Fluides Organis\'es, FRE 2844 du CNRS, Coll\`ege de France, 11 Place Marcelin Berthelot, 75231 Paris Cedex 05, France
}

\author{Pascal Damman}
\author{S\'everine Sclavons}
\author{Sylvain Gabriele}
\altaffiliation[Also at ]{MateriaNova Research Center, Mons, Belgium}
\affiliation{
Laboratoire de Physicochimie des Polym\`eres, Universit\'e de Mons Hainaut, 20 Place du Parc, Mons B-7000, Belgium
}

\author{G{\"u}nter Reiter}
\author{Moustafa Hamieh}
\altaffiliation[Also at ]{Laboratoire de Chimie Analytique, Mat\'eriaux, surfaces et Interfaces, D\'epartement de Chimie, Facult\'e des Sciences I, Universit\'e Libanaise, Hadeth, Beyrouth, Liban}
\affiliation{
Institut de Chimie des Surfaces et Interfaces, CNRS-UHA, 15 rue Jean Starcky, B. P. 2488, 68057 Mulhouse cedex, France}

\date{\today}

\begin{abstract}
The study of the dewetting of very thin polymer films has
recently revealed many unexpected features (\textit{e.g.}
unusual rim morphologies and front velocities) which have
been the focus of several theoretical models.
%have sparked of many theoretical studies.
Surprisingly, the most striking feature of all, that is a decrease of the rim width with time, %during the second part of the dewetting process,
%which is a first time in dewetting science, 
have not yet been explained.
In the present letter, we show how the combined effects of a non-linear friction between the film and the substrate, and the presence of residual stresses within the film, result in the presence of a maximum 
in the time evolution of the rim width.
%of the rim width, followed by a decrease of the rim width with time. 
In addition, we show how the introduction of a non-linear friction
%in the description of the dewetting of thin polymer films was found to
can also simply explain the rapid decrease of the dewetting velocity with time observed experimentally.
%which was not fully understood until now.
\end{abstract}

\maketitle

Since the use of nano-devices %of systems with very small dimensions
is nowadays getting more and more important, the stability of
%of nano-structures like 
thin polymer films 
%(as well as the influence of confinement on polymers, 
has become a major scientific and technologic research area \cite{speissue}.
In particular, the dewetting of thin polystyrene (PS) films 
%close to the glass transition 
has been extensively investigated over the last decade \cite{bucknall}. A great interest has been payed to the unusual rim morphology \cite{saulnier2002, shenoy2002, herminghaus} first studied by Reiter \cite{reiter2001}. A detailed experimental study of the dewetting dynamic has been published two years ago \cite{damman2003}, showing a sharp decrease of the dewetting velocity with time ($V \sim t^{-1}$).\\
\indent These observed features have been recently explained theoretically in \cite{vilmin2005}. The principal ingredients allowing to understand the observed asymmetric shape of the rim are the weak friction of the film onto the substrate (grafted or absorbed liquid polydimethylsiloxane (PDMS) mono-layer), allowing a plug-flow description, and the ineffectiveness of the surface tension of the liquid at this stage of the dewetting process. The friction dumps the sliding velocity within the film over a distance $\Delta = \sqrt{\eta h_{0}/\zeta_{0}}$ (where $\eta$ is the viscosity of the liquid, assumed to be Newtonian, $h_{0}$ is the initial thickness of the film, and $\zeta_{0}$ is the friction coefficient between the film and the substrate) \cite{brochard97}, leading to an asymmetric rim of width $\sim \Delta$, whereas the height of the rim $H$ increases progressively. Omitting inertia, the dewetting dynamic can be simply solved by balancing the capillary power by the dissipation due to friction: $|S|V \simeq \zeta_{0} \Delta V^{2}$ (where $S$ is the spreading parameter \cite{PGGFBDQ}, assumed to be negative). For a Newtonian liquid, this balance gives a constant dewetting velocity $V =|S|/\zeta_{0} \Delta$ \cite{brochard97}.
%%%%%%%%%%
\begin{figure}
\includegraphics[clip=true, bb=2cm 8.5cm 18.5cm 18cm, width = 7.0cm]{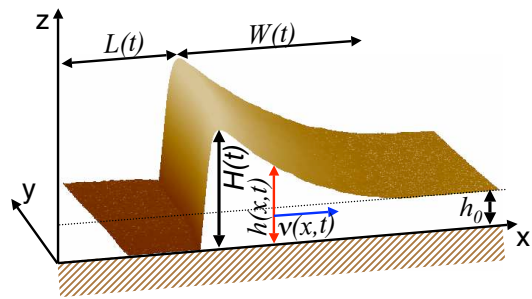}
\caption{\label{fig1} Film geometry : $h(r, t)$ is the profile of the film, $h_{0}$ is the initial height of the film, $H(t)$ is the height of the front, $L(t)$ is the dewetted distance, $W(t)$ is the width of the rim, and $v(x, t)$ is the velocity of the film.}
\end{figure}
%%%%%%%%%%
The decrease with time of the velocity observed experimentally is explained by the viscoelastic properties of the polymer film, {\it i.e.} by the fact that the liquid behaves like a low viscosity ($\eta_{0}$) Newtonian liquid at very short times ($t < \tau_{0} = \eta_{0}/G$), like an elastic solid (of elastic modulus $G$) in an intermediate time regime ($\tau_{0} < t < \tau_{1} = \eta_{1}/G$), and like a very viscous Newtonian liquid (with a viscosity $\eta_{1} \gg \eta_{0}$, where $\eta_{1}$ is related to the molecular weight of the polymer) at long times ($t > \tau_{1}$) \cite{bird}. During the short and long times regimes \cite{longertimes}, the film dewetts like a Newtonian liquid, with the respective  constant velocities $V_{0} =|S|/\zeta_{0} \Delta_{0}$ ($t < \tau_{0}$, $\Delta_{0} = \sqrt{\eta_{0} h_{0}/\zeta_{0}}$) and $V_{1} =|S|/\zeta_{0} \Delta_{1}$ ($t > \tau_{1}$, $\Delta_{1} = \sqrt{\eta_{1} h_{0}/\zeta_{0}} \gg \Delta_{0}$). In the intermediate  regime (which concerns an important part of the experiments, for which the temperature is close to the glass transition temperature) the increase of the height of the rim is slowed down by elastic forces, resulting in a linear relation between the width of the rim $W$ and the dewetted distance $L$. The energy balance thus reads $|S|V \, \simeq \, \zeta_{0} h_{0}GLV^{2}/|S|$ which gives a decrease of the velocity like $t^{-\frac{1}{2}}$. The more rapid decrease ($V \sim t^{-1}$) observed experimentally \cite{damman2003} is obtained by including a positive horizontal stress $\sigma_{0}$ within the film at the beginning of the dewetting process; the existence of $\sigma_{0}$ is attributed to the spin-coating of the PS solution, which is followed by a fast evaporation of the solvent, letting the PS molecules in a non-equilibrium frozen-in state \cite{reiterpgg}.\\
\indent Until now, little attention has been paid to the time evolution of the rim width reported in \cite{damman2003}, though it is most unexpected. Indeed, after logarithmically increasing, the rim width reaches a maximum, and then slowly goes down.
%in an approximately logarithmic way. 
One can show that such a decrease of the rim width with time is impossible in a classical dewetting situation where surface tension rounds the rim. In the theoretical model described above \cite{vilmin2005}, despite the fact that surface tension is ineffective, the rim width of a viscoelastic film increases with time during the elastic regime ($W \sim L$), while reaching a constant value in the long time Newtonian regime: $W = \Delta_{1} = \sqrt{\eta_{1} h_{0}/\zeta_{0}}$, therefore showing no decrease with time of the rim width. However, at the sight of the latter expression of $W$, we can anticipate that $W$ could decrease at the condition that the friction parameter $\zeta_{0}$ increases with time.\\
%, and where the rim width is directly related to the dewetted distance by volume conservation ({\it i.e.} $WH \sim W^{2} \sim Lh_{0}$). 
%At first sight such a decrease is conceivable in the theoretical model described above \cite{vilmin2005}, where the surface tension is ineffective and where the rim width and the rim height are thus not linearly linked. A decrease of the rim width would, then, mean a displacement backward of the contact line.
%However, in the theoretical results we reported from \cite{vilmin2005}, for a viscoelastic liquid, the rim width increases with time during the elastic regime ($W \sim L$), while reaching a constant value in the long time Newtonian regime: $W = \Delta_{1} = \sqrt{\eta_{1} h_{0}/\zeta_{0}}$. At the sight of this latter expression, we can anticipate that $W$ could decrease at the condition that the friction parameter $\zeta_{0}$ increases with time.\\
\indent A non-constant friction parameter is possible in the case where the friction force does not increase linearly with the sliding velocity. These cases are quite common in polymer physics. The friction of a PDMS elastomer on grafted or absorbed PDMS surfaces has been shown to increase only very slowly with the velocity: Casoli {\it et al.} found a friction force increasing proportionally to $V^{\frac{1}{3}}$ on absorbed brushes, and a friction force increasing only like $V^{\frac{1}{6}}$ on dense grafted brushes, for sliding velocities being between $10 \mu m.s^{-1}$ and $5$ $mm.s^{-1}$ \cite{casoli2001}. More recently Bureau {\it et al.} showed a $V^{\frac{1}{5}}$ dependence of the friction force between an elastomer and a grafted brush, for sliding velocities ranging from $300 \mu m.s^{-1}$ down to $0.01 \mu m.s^{-1}$ \cite{bureau2004}, which is the velocity range of the dewetting experiments.\\
\indent A general expression of the friction force by surface unit is \cite{friction}:
%%%%%%%%%%
\begin{equation}
f_{r} \, = \, \zeta_{0} V_{\alpha}\left(\frac{v}{V_{\alpha}}\right)^{1-\alpha}
\label{eq2}
\end{equation}
%%%%%%%%%%
%for $v > V_{\alpha}$, 
where $\alpha$ is a shear-thinning exponent smaller than one. According to the reported friction experiments, $\alpha$ could range between $2/3$ and $5/6$. The effective friction coefficient is then a decreasing function of $v$: $\zeta = \zeta_{0} \left(V_{\alpha}/v\right)^{\alpha}$, where $V_{\alpha}$ is the velocity below which the friction coefficient remains equal to $\zeta_{0}$ correspondingly to a linear friction. Hereafter, we assume $V_{\alpha}$ to be small enough for us to omit the linear friction regime.
%For $v < V_{\alpha}$, the friction coefficient remains equal to $\zeta_{0}$, correspondingly to a linear friction. The velocity $V_{\alpha}$ is assumed to be smaller than the observed dewetting velocities, which allows us to omit the linear friction regime.
For a Newtonian fluid, such a friction force would bring, as an equilibrium relation between friction and viscous forces within a film dewetting toward the $x$ direction:
%%%%%%%%%%
\begin{equation}
\zeta_{0} V_{\alpha}\left(\frac{v}{V_{\alpha}}\right)^{1-\alpha} =  \, \eta_{0} \,\frac{\partial}{\partial x} \left( h \frac{\partial v}{\partial x}\right)
\label{eq3}
\end{equation}
%%%%%%%%%%
As long as $h$ remains of the order of $h_{0}$, the solution of Eq.(\ref{eq3}), is simply given by
%%%%%%%%%%
\begin{equation}
v(x, t) \, = \, V_{0\alpha} \left(1 - \frac{x-V_{0\alpha}t}{\Delta_{0\alpha}}\right)^{\frac{2}{\alpha}}
\label{eq4}
\end{equation}
%%%%%%%%%% \left((2-\alpha}{2}\right)^{\frac{1}{2-\alpha}} V_{0}^{\frac{2}{2-\alpha}} V_{\alpha}^{-\frac{\alpha}{2-\alpha}}
when $x-V_{0\alpha}t < \Delta_{0\alpha}$, and $v(x, t) = 0$ elsewhere. Here, the dewetting velocity $V_{0\alpha} =  \left(\left(\frac{2-\alpha}{2}\right)V_{0}^{\,2}/V_{\alpha}^{\,\,\alpha}\right)^{\frac{1}{2-\alpha}}$ is constant in time, as well as the rim width $\Delta_{0\alpha} = \frac{2}{\alpha}\left(\left(\frac{2-\alpha}{2}\right) V_{0}^{\,\alpha}/V_{\alpha}^{\,\,\alpha}\right)^{\frac{1}{2-\alpha}} \Delta_{0}$ \cite{rim}. These analytical results have been confirmed by numerically solving Eq.(\ref{eq3}) combined with the volume conservation equation and the condition of stress continuity at the front of the film.\\
%%%%%%%%%%
\begin{figure}
\includegraphics[clip=true, bb=6cm 1.5cm 24cm 17cm, width = 6.5cm, height = 5cm]{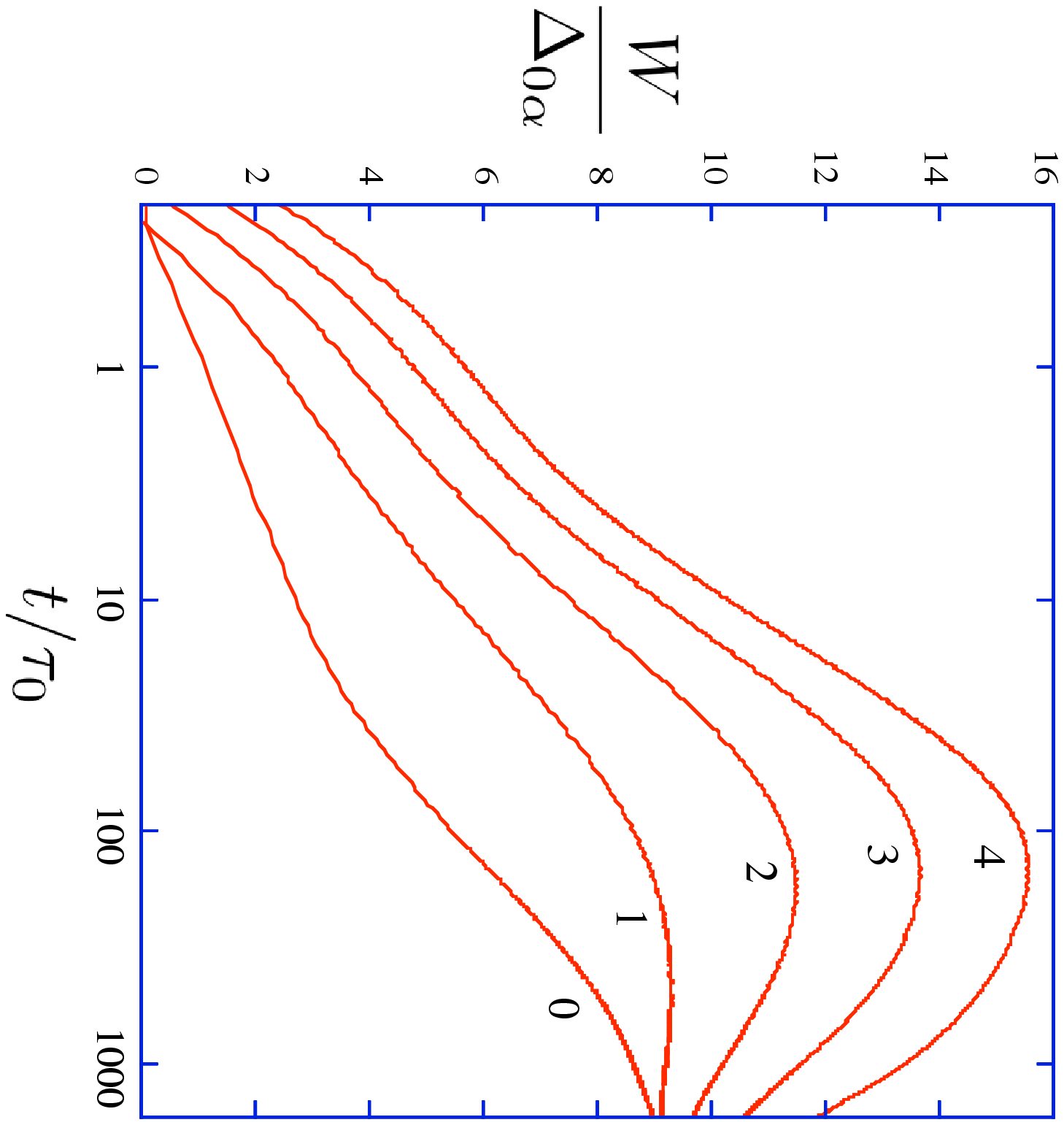}
\caption{\label{fig2} Numerical calculations of the evolution of the rim width for $h_{0}\sigma_{0}/|S| = 0$, $1$, $2$, $3$, and $4$, with a friction exponent $\alpha = 2/3$, and a ratio $\tau_{1}/\tau_{0} = 100$.}
\includegraphics[clip=true, bb=0.5cm 0.5cm 11cm 8cm, width = 6.5cm]{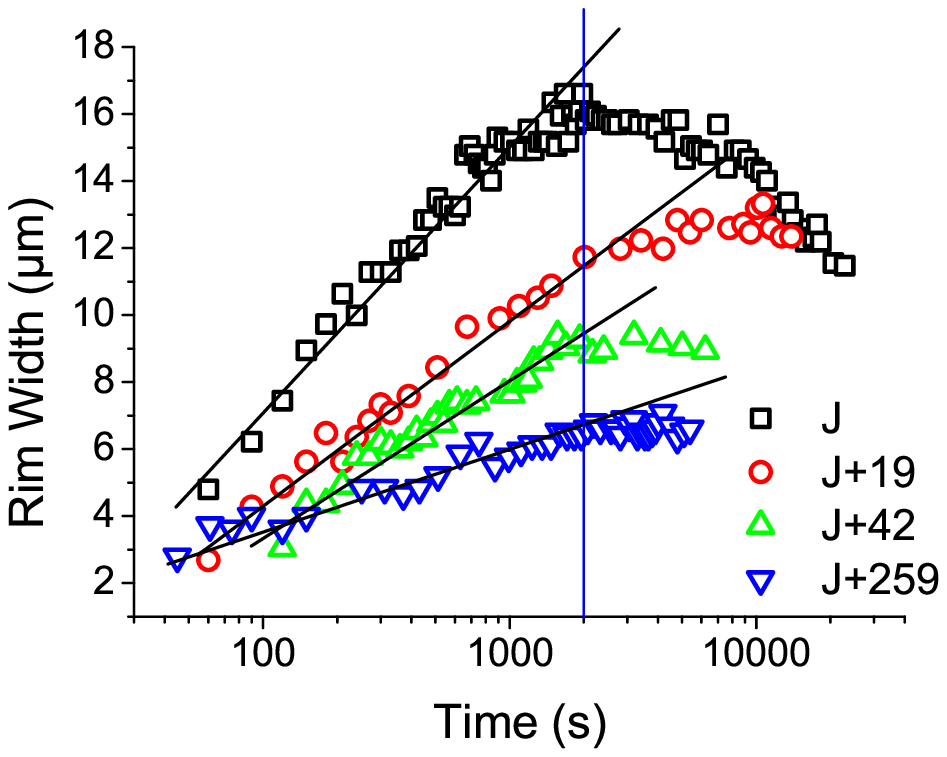}
\caption{\label{fig3} Width of the rim as a function of dewetting time for films ($h_{0} = 57 nm$, $M_{w} = 233 kg/mol$, and $T=125^{o}C$) aged at room temperature for days indicated in the figure.}
\end{figure}
%%%%%%%%%%
\indent As expected, when the friction force is a non-linear function of the velocity, the rim width is a function of the amplitude of the dewetting driving forces ({\it i.e.} the capillary forces in usual cases: $\Delta_{0\alpha} \sim |S|^{\frac{\alpha}{2-\alpha}}$). A decrease of the rim width can thus be expected if the driving forces decrease during the dewetting process.
That is, in fact, what happens when a viscoelastic film presents residual stresses at the onset of the dewetting. Indeed, the residual positive stresses are part of the driving forces, but they decrease with time, as the film undertakes internal relaxation toward an equilibrated state \cite{reiterpgg, vilmin2005}.\\
\indent An equilibrated viscoelastic film would start ($t < \tau_{0}$) dewetting with a velocity $V_{0\alpha} \simeq (V_{0}/V_{\alpha}^{\, \frac{\alpha}{2}})^{\frac{2}{2-\alpha}}$, and a rim width $\Delta_{0\alpha} \simeq \left(V_{0}/V_{\alpha}\right)^{\frac{\alpha}{2-\alpha}} \Delta_{0}$, and then ($\tau_{0} < t < \tau_{1}$) slow down to $V_{1\alpha} \simeq (V_{1}/V_{\alpha}^{\, \frac{\alpha}{2}})^{\frac{2}{2-\alpha}}$, ending with a rim width $\Delta_{1\alpha} \simeq \left(V_{1}/V_{\alpha}\right)^{\frac{\alpha}{2-\alpha}} \Delta_{1}$ at long times ($\tau_{1} < t$). When a horizontal positive stress $\sigma_{0}$ is initially present, the film starts dewetting with a velocity $V_{i} = V_{0\alpha}\left(1+h_{0}\sigma_{0}/|S|\right)^{\frac{2}{2-\alpha}}$ and a rim width $\Delta_{i} = \Delta_{0\alpha}\left(1+h_{0}\sigma_{0}/|S|\right)^{\frac{\alpha}{2-\alpha}}$. During the elastic regime ($\tau_{0} < t < \tau_{1}$), the residual stress has not yet relaxed much, and the rim width increases up to its maximum value:
%%%%%%%%%%
\begin{equation}
\Delta_{m} \simeq \Delta_{1\alpha}\left(1+\frac{h_{0}\sigma_{0}}{|S|}\right)^{\frac{\alpha}{2-\alpha}}\label{eq5}
\end{equation}
%%%%%%%%%%
Around the characteristic time $\tau_{1}$, the residual stress has decreased significantly, leaving the capillary forces as the only driving forces. Consequently, the rim width decreases down to $\Delta_{1\alpha}$ simultaneously with the increase of the dewetting distance. Once again, we confirmed these results by numerical resolutions of the equations of the flow (see Fig. (\ref{fig2})).
 We notice that $\Delta_{m}$ is higher than $\Delta_{1\alpha}$ only if the exponent $\alpha$ is not nil. The combination of non-linear friction together with residual stresses is thus necessary to explain the decrease of the rim width with time. Interestingly, the residual stresses $\sigma_{0}$ can come off from the relation between the initial dewetting velocity $V_{i}$ and the maximum rim width $\Delta_{m}$:
 %%%%%%%%%%
\begin{equation}
V_{i} = V_{\alpha}\left(\frac{\zeta_{0}}{h_{0}}\right)^{\frac{1}{\alpha}}\left(\frac{\eta_{1}^{\, 2(1-\alpha)}}{\eta_{0}}\right)^{\frac{1}{2-\alpha}}\Delta_{m}^{\, \, \frac{2}{\alpha}}\label{eq5bis}
\end{equation}
%%%%%%%%%%
The exponent $\alpha$ could thus be deduced from a Log-Log plot of $V_{i}$ as a function of $\Delta_{m}$, for dewetting experiments with various values of the residual stress. Furthermore, we notice a weak dependance of $V_{i}/\Delta_{m}^{\,\,\frac{2}{\alpha}}$ on the viscosity $\eta_{1}$, and thus on the molecular weight of the polymer, when $\alpha$ is close to unity. Then, several Log-Log plots, corresponding to different molecular weights, can be superposed in order to measure $\alpha$.\\
%Furthermore, we can notice that the ratio $V_{\sigma_{0}\alpha}/\Delta_{max}^{\frac{2}{\alpha}}$ depends only very weakly on the viscosity $\eta_{1}$, and thus on the molecular weight of the polymer, when the exponent $\alpha$ is close to unity. Therefore, several $Log-Log$ plots of $V_{\sigma_{0}\alpha}$ as a function of $\Delta_{max}$, corresponding to experiments using different molecular weight, can be superposed in order to measure $\alpha$.
%%%%%%%%%%
\begin{figure}
\includegraphics[clip=true, bb=0.5cm 0.5cm 11cm 8cm, width = 6.5cm]{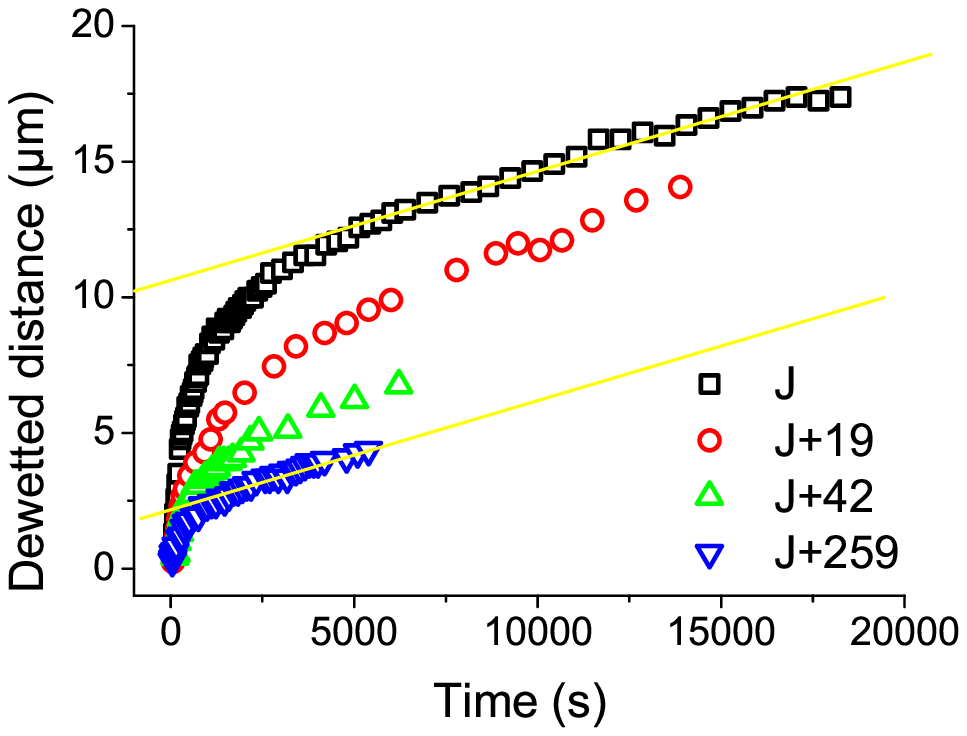}
\caption{\label{fig4} Dewetted distance as a function of dewetting time for films ($h_{0} = 57 nm$, $M_{w} = 233 kg/mol$, and $T=125^{o}C$) aged at room temperature for days indicated in the figure.}
\includegraphics[clip=true, bb=0.5cm 0.5cm 11.5cm 8.5cm, width = 6.5cm]{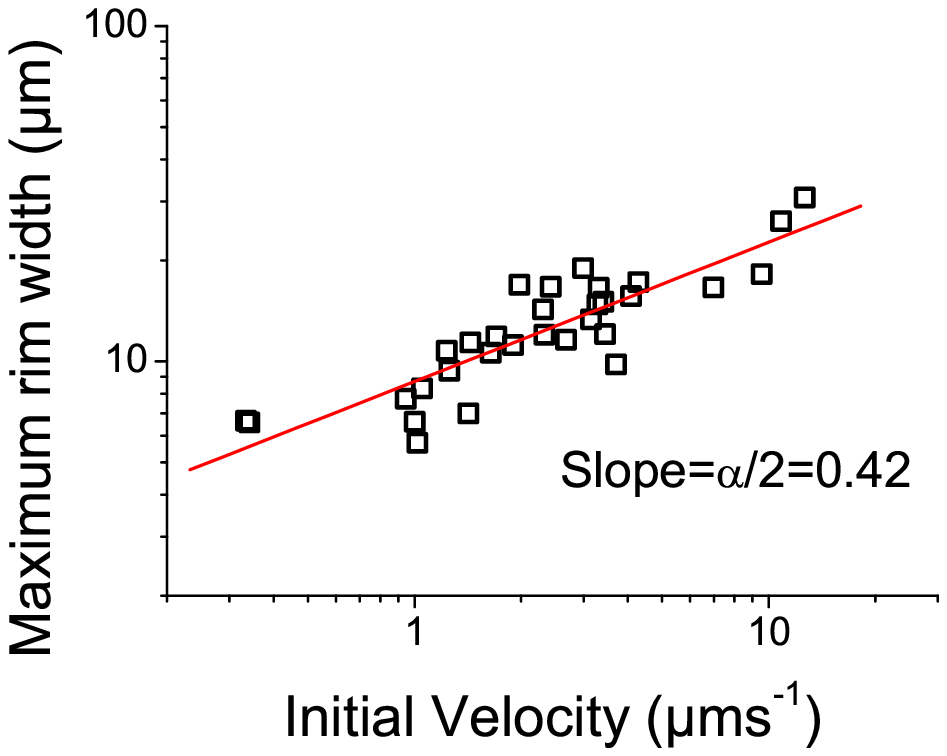}
\caption{\label{fig5} Evolution of the maximum rim with as a function of the initial dewetting velocity for several aging times and PS molecular weights ranging from $35K$ to $600K$.}
\includegraphics[clip=true, bb=6cm 1.5cm 24cm 17cm, width = 6.5cm, height = 5cm]{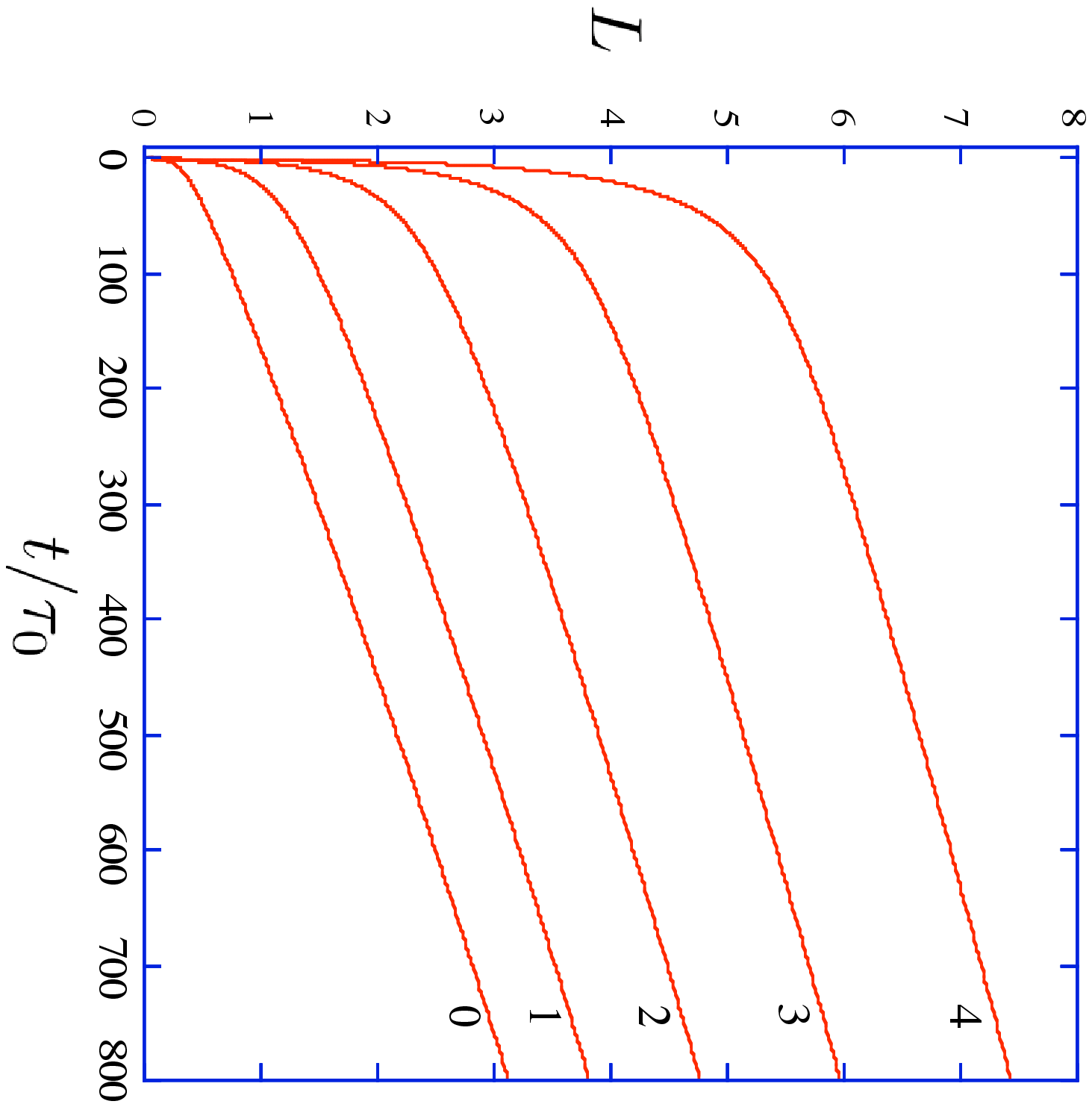}
\caption{\label{fig6} Numerical calculations of the evolution of the dewetted distance for $h_{0}\sigma_{0}/|S| = 0$, $1$, $2$, $3$, and $4$, with the friction exponent $\alpha = 2/3$, and a ratio $\tau_{1}/\tau_{0} = 100$.}
\end{figure}
%%%%%%%%%%
\indent At the same time, we have conducted new experiments on the aging of thin PS films deposited onto a PDMS mono-layer. The consequences of up to three hundred days of aging, just below the glass transition, on the dewetting have been systematically investigated. We observed a slowing down of the dewetting, as well as a decrease of the maximum rim width, when increasing the aging time (see Fig. (\ref{fig3}) and (\ref{fig4})). %, but no consequences on the position in time of this maximum 
Eventually, at the longest aging time, no maximum rim width was observed. We interpreted these results as a manifestation of the very slow relaxation of the residual stresses within the film with aging time. The Log-Log plot of the maximum rim with as a function of the initial dewetting velocity for several aging times and several PS molecular weight revealed a friction exponent $\alpha = 0.84 \pm 0.1$ (see Fig. (\ref{fig5})). The time corresponding to the maximum rim width should correspond to the characteristic relaxation time of elastic constraints $\tau_{1}$ within the film. We notice that this time is significantly shorter than the corresponding relaxation time in bulk ({\it i.e.} the bulk reptation time of the polymer chains), which is consistent with the fact that the entanglement length should be longer in thin films \cite{si2005}. Other structural relaxation mechanisms related to the proximity of the glass transition might also be involved.\\
\indent Another interesting result, which reinforces the implication of non-linear friction in dewetting, is that, even-though the initial dewetting velocity $V_{i}$ decreases with the aging time, the power-law evolution of the velocity during the dewetting process remains the same ($V \simeq t^{-1})$. This observation is in contradiction with the predictions of the theoretical model developed  in \cite{vilmin2005}, where a $t^{-1}$ law was predicted for high residual stresses, but where the decrease of the velocity should tend toward a $t^{-\frac{1}{2}}$ law for low residual stresses. On the other hand, this observation is coherent with the model we developed here. Indeed, the dewetting velocity of an equilibrated viscoelastic film goes from $V_{0\alpha}$ down to $V_{1\alpha}$ during the elastic regime. The power law of this decrease can be obtained from the energy balance, which reads $|S|V \simeq \zeta_{0}h_{0}GV_{\alpha}^{\,\,\alpha}LV^{2-\alpha}/|S|$. The later relation gives:
%%%%%%%%%%
%\begin{equation}
%|S|V = \zeta_{0}\frac{h_{0}G}{|S|}V_{\alpha}^{\,\,\alpha}LV^{2-\alpha}
%\label{eq6}
%\end{equation}
%%%%%%%%%%
%%%%%%%%%%
\begin{equation}
V = V_{1\alpha} \left(\frac{t}{\tau_{1}}\right)^{-\frac{1}{2-\alpha}}
\label{eq7}
\end{equation}
%%%%%%%%%%
for $\tau_{0} < t < \tau_{1}$, which is not far from $V \sim t^{-1}$ if $\alpha$ is close to unity, as deduced from the plot of $W_{m}$ as a function of $V_{i}$ (see Fig. (\ref{fig5})).
 %for $\frac{2}{3} < \alpha < \frac{5}{6}$.
 Thus, the dewetting velocity has always a power law close to $t^{-1}$, which is made more rapid around the relaxation time $\tau_{1}$ by the presence of residual stresses. This latter result is in complete agreement with our experiments (see also \cite{damman2003, reiter2003}). We have also confirmed these analytical prediction by numerical solutions (see Fig. (\ref{fig6}) and (\ref{fig7})).\\
%%%%%%%%%%
\begin{figure}
\includegraphics[clip=true, bb=6cm 1.5cm 24cm 17cm, width = 6.5cm, height = 5cm]{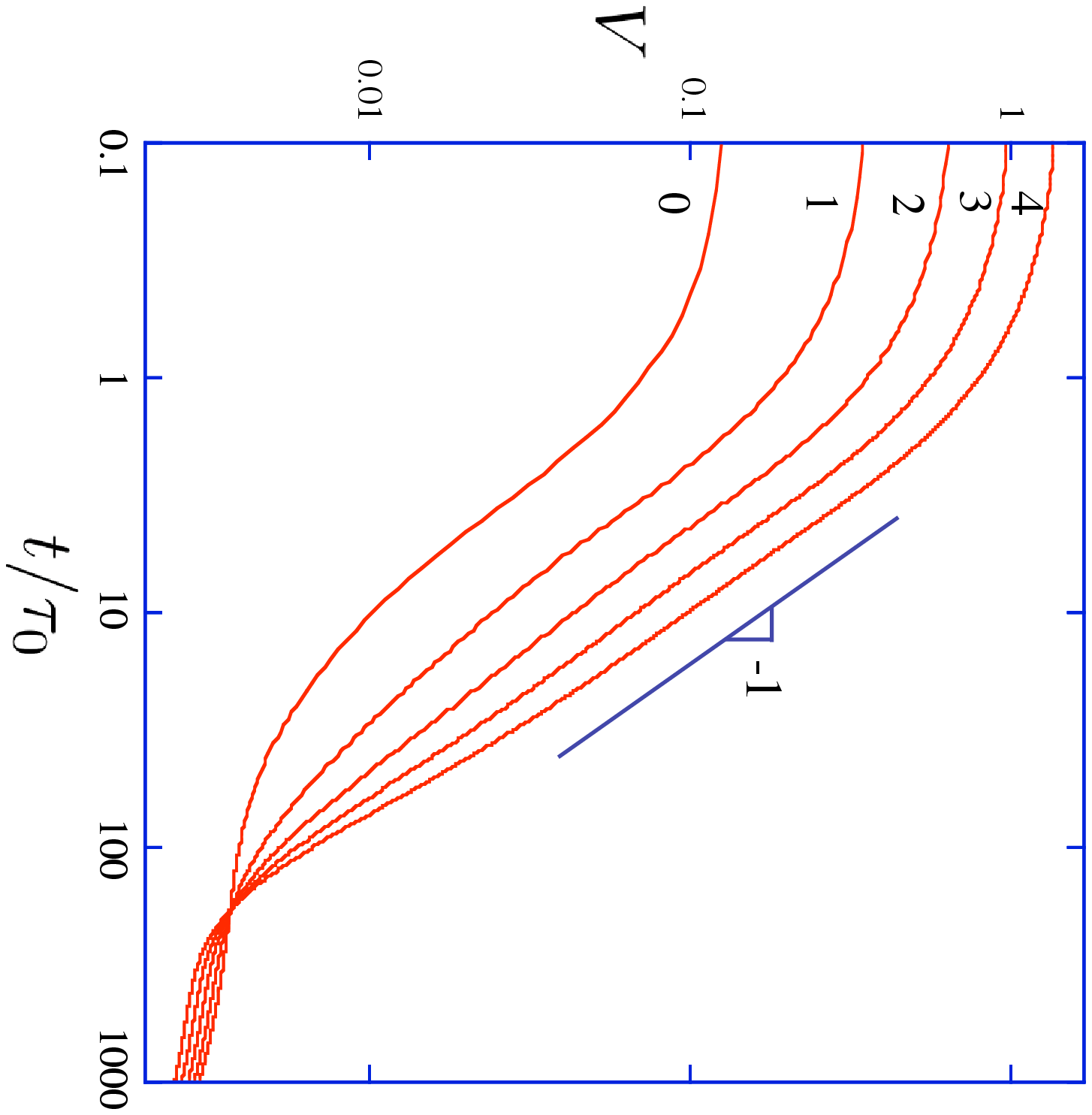}
\caption{\label{fig7} Numerical calculations of the evolution of the dewetting velocity for $h_{0}\sigma_{0}/|S| = 0$, $1$, $2$, $3$, and $4$, with the friction exponent $\alpha = 2/3$, and a ratio $\tau_{1}/\tau_{0} = 100$.}
\end{figure}
%%%%%%%%%%
\indent In conclusion, the non-linearity of the friction force between the substrate and the film, together with the initial presence of residual stress within the dewetting liquid, are necessary to explain the observed decrease of the rim width with time. Coincidentally, the non-linear friction leads to a rapid decrease of the dewetting velocity of a viscoelastic film, which fits the experiments better than the velocity decrease predicted by the formers models. The good agreement of the predictions on both the rim width evolution and the dewetting velocity makes us confident about the validity of our model. Finally, the Log-Log plot of the initial dewetting velocities as a function of the maximum rim width allows to determine the exponent $\alpha$ characterizing the friction between the film and the substrate. Our study clarifies the role
played by the friction and the residual stresses, both ingredients that
may be of great importance in other polymer thin film situations.
A major challenge for the future would be to understand the processes involved in the relaxation of the residual stresses below the glass transition.

\acknowledgments
Pascal Damman is a Research Associate of the Belgian National Funds for Scientific Research (FNRS). This work was supported by the FNRS, the Walloon Region and the European Social Fund.


\begin{thebibliography}{9}

\bibitem{speissue}
G. Reiter, J. Forrest, Special Issue on Properties of Thin Polymer Films. Eur. Phys. J. E {\bf 8}, 101 (2002)

\bibitem{bucknall}
D. G. Bucknall, Progress in Materials Science {\bf 49}, 713 (2004)

\bibitem{saulnier2002}
F. Saulnier, E. Rapha\"{e}l, and P.-G. de Gennes, Phys. Rev. Lett.
{\bf 88}, 196101 (2002); F. Saulnier, E. Rapha\"{e}l, and P.-G. de Gennes, Phys. Rev. E
{\bf 66}, 061607-(1-12) (2002)

\bibitem{shenoy2002}
V. Shenoy and A. Sharma, Phys. Rev. Lett. {\bf 88}, 236101 (2002)

\bibitem{herminghaus}
S. Herminghaus, R. Seemann, and K. Jacobs, Phys. Rev. Lett. {\bf 89}, 056101 (2002)

\bibitem{reiter2001}
G. Reiter, Phys. Rev. Lett. {\bf 87}, 186101 (2001)

\bibitem{damman2003}
P. Damman, N. Baudelet, and G. Reiter, Phys. Rev. Lett. {\bf 91}, 216101 (2003)

\bibitem{vilmin2005}
T. Vilmin, and E. Rapha\"{e}l, cond-mat/0502228 (2005).

\bibitem{brochard97}
F. Brochard-Wyart, G. Debr\'egeas, R. Fondecave, and P. Martin, Macromolecules {\bf 30}, 1211 (1997)

\bibitem{PGGFBDQ}
P.-G. de Gennes, F. Brochard-Wyart and D. Qu\' er\'e,
Capillarity and Wetting Phenomena: Drops, Bubbles, Pearls, Waves
(Springer, 2003).

\bibitem{bird}
R. B. Bird, R. C. Armstrong, and O. Hassager, {\it Dynamics of polymeric liquids} (John Wiley \& Sons, volume 1, 1977)

\bibitem{longertimes}
Eventually, at even longer times, surface tension will round the rim, leading to a $t^{-\frac{1}{3}}$ decrease of the velocity (see \cite{brochard97, vilmin2005}).

\bibitem{reiterpgg}
G. Reiter, and P.-G. de Gennes, Eur. Phys. J. E. {\bf 6}, 25
(2001)

\bibitem{casoli2001}
A. Casoli, M. Brendl\'e, J. Schultz, P. Auroy, and G. Reiter, Langmuir {\bf 17}, 388 (2001)

\bibitem{bureau2004}
L. Bureau, and L. L\' eger, Langmuir {\bf 20}, 4523 (2004)

\bibitem{friction}
Even if almost no interpenetration is expected between the PS film and the PDMS layer, the non-linear relation between friction and velocity could be a consequence of the rheologic response of the brush to a shear strain, \textit{i.e.} shear thinning.
%The weak exponents characterizing these relations between the friction force and the sliding velocity are the results of both the variation of the interpenetration between the elastomer and the brush, and the rheologic response of the brush to a shear strain. In the case of a PS film sliding over a PDMS brush, almost no interpenetration is expected, and thus the relation between the friction and the velocity is mostly related to the rheologic properties of the brush.

\bibitem{rim}
Note that when $\alpha$ tends toward zero, $V_{0\alpha}$ tends toward $V_{0}$, and $\Delta_{0\alpha}$ scales like $\Delta_{0}$ but diverges. This is a simple consequence of the transition from a finite power law shape to an exponential shape of the rim.
%This is because when the friction is linear the rim has not a finite power law shape, but an exponential shape, with the characteristic decreasing length $\Delta_{0}$.

\bibitem{si2005}
L. Si, M. V. Massa, K. Dalnoki-Veress, H. R. Brown, and R. A. L. Jones, Phys. Rev. Lett. {\bf 94}, 127801 (2005)

\bibitem{reiter2003}
G. Reiter, M. Sferrazza, and P. Damman, Eur. Phys. J. E {\bf 12}, 133 (2003)

%\bibitem{oron}
%A. Oron, S. H. Davis, and S. G. Bankoff, Rev. Mod. Phys. {\bf 69}, 931 (1997)

%\bibitem{revues}
%G. Reiter, Science {\bf 282}, 888 (1998) 

%\bibitem{Dalnoki1999}
%K. Dalnoki-Veress, B.G. Nickel, C. Roth, and J.R. Dutcher, Phys. Rev. E
%{\bf 59}, 2153 (1999)

%\bibitem{redon94}
%C. Redon, J. B. Brzoska, and F. Brochard-Wyart, Macromolecules {\bf 27}, 468
%(1994)

%\bibitem{masson}
%J.-L. Masson and P. F. Green, Phys. Rev. E {\bf 65}, 031806 (2002)

%\bibitem{glass}
%Note that at even shorter times, the fluid behaves like a glassy material; see {\it{e.g.}} 
%.P-G. de Gennes, Scaling Concepts in Polymer Physics (Cornell University Press, 1979)



%\bibitem{forrest}
%J. A. Forrest, K. Dalnoki-Veress, and J. R. Dutcher, Phys. Rev. E
%{\bf 56}, 5705 (1997)

%\bibitem{taylor}
%G. I. Taylor, Proc. R. Soc. London, Ser. A. {\bf 253}, 313 (1959)

%\bibitem{culick}
%F. E. C. Culick, J. Appl. Phys. {\bf 31}, 1128 (1960)

%\bibitem{keller1}
%J. B. Keller, Phys. Fluids {\bf 26}, 3451
%(1983)

%\bibitem{mysels}
%W. R. Mc Entee, and K. J. Mysels, J. Phys. Chem. {\bf 73},
%3018/3028 (1969)

%\bibitem{keller}
%J. B. Keller, A. King, and L. Ting, Phys. Fluids
%{\bf 7}, 226 (1995)

%\bibitem{debregeas}
%G. Debr\'egeas, P. Martin, and F. Brochard-Wyart, Phys. Rev. Lett.
%{\bf 75}, 3886 (1995); \\ G. Debr\'egeas, P.-G. de Gennes, and F.
%Brochard-Wyart, Science {\bf 279}, 1704 (1998)

%\bibitem{brenner}
%M. P. Brenner and D.
%Gueyffier, Phys. Fluids {\bf 11}, 737 (1999)

%\bibitem{koplik}
%J. Koplik, and J. R. Banavar, Phys. Rev. Lett. {\bf 84},
%4401 (2000)

%\bibitem{liu}
%Kinetics of growth of dry patches (during the early stages of
%dewetting process of microscopically thin polymer films cast on
%a solid surface) are simulated by molecular dynamics in: H. Liu, A.
%Bhattacharya, and A. Chakrabarti, J. Chem. Phys. {\bf 109}, 8607
%(1998)

%\bibitem{herminghaus}
%For somewhat related studies, see: R. Seemann, S. Herminghaus, and
%K. Jacobs, Phys. Rev. Lett. {\bf 87}, 196101 (2001)

%\bibitem{condmat}
%S. Herminghaus, R. Seemann, and K. Jacobs, condmat/0201193

%\bibitem{dalnoki}
%K. Dalnoki-Veress {\it et al.}, Phys. Rev. E, {\bf 59}, 2153
%(1999)

%\bibitem{rem2}
%P.-G. de Gennes, C. R. Acad. Sci. {\bf 288B}, 219 (1979)

%\bibitem{Herminghaus2001}
%For a related study, see, S. Herminghaus, R. Seemann, and K. Jacobs,  
%Euro. Phys. J. E {\bf 87}, 196101 (2001)



%\bibitem{christensen}
%R. M. Christensen, {\it Theory of viscoelasticity, an
%introduction} (Academic Press, 2nd edition, 1982), p.53

%\bibitem{redon}
%C. Redon, F. Brochard-Wyart, and F. Rondelez, Phys. Rev.
%Lett. {\bf 66}, 715 (1991)

%\bibitem{gunterprec}
%G. Reiter, Phys. Rev. Lett. {\bf 68}, 75 (1992)

%\bibitem{safran}
%S. A. Safran and J. Klein, J. Phys. II France {\bf 3}, 749 (1993)

%\bibitem{francoiseetpgg}
%On the wetting and slippage of polymer films on solid surfaces,
%see: F. Brochard-Wyart {\it et al.}, Langmuir {\bf 10}, 1566
%(1994); C. Redon, J. B. Brozoska, and F. Brochard-Wyart, Macromol.
%{\bf 27}, 468 (1994)

\end{thebibliography}
\end{document}